\documentclass[epj]{webofc}
\usepackage[varg]{txfonts}   % Web of Conferences font
\usepackage{graphicx}
\usepackage{epsfig}
\woctitle{MESON2014 - the 13$^\textrm{th}$ International Workshop on Meson Production, Properties and Interaction}
\begin{document}

\selectlanguage{english}
\title{Search for the $^{4}\hspace{-0.03cm}\mbox{He}$-$\eta$ bound state in $dd\rightarrow(^{4}\hspace{-0.03cm}\mbox{He}$-$\eta)_{bound}\rightarrow$ $^{3}\hspace{-0.03cm}\mbox{He} n \pi{}^{0}$ and $dd\rightarrow(^{4}\hspace{-0.03cm}\mbox{He}$-$\eta)_{bound}\rightarrow$ $^{3}\hspace{-0.03cm}\mbox{He} p \pi{}^{-}$ reactions with the WASA-at-COSY facility}

% insert email only for speaker/presenter
\author{M. Skurzok\inst{1}\fnmsep\thanks{\email{magdalena.skurzok@uj.edu.pl}} \and W. Krzemien \inst{1} \and P. Moskal \inst{1,2}
\\for the WASA-at-COSY Collaboration
}

\institute{M. Smoluchowski Institute of Physics, Jagiellonian University, Cracow, Poland \and Institut fur Kernphysik, Forschungszentrum J\"ulich, Germany}

\abstract{
%In 1986 Haider and Liu postulated a new kind of exotic nuclear matter - $\eta$-mesic nuclei in which the $\eta$ meson is bound in a nucleus by means of the strong interaction \cite{HaiderLiu1}. However, till now there is no clear experimental evidence confirmed empirically its existence.
In November 2010, the search for the $^{4}\hspace{-0.03cm}\mbox{He}$-$\eta$ bound state was performed with high statistics and high acceptance with the WASA-at-COSY facility using a ramped beam technique. The signature of $\eta$ - mesic nuclei is searched for in the measured excitation functions for the two reaction channels: $dd\rightarrow$ $^{3}\hspace{-0.03cm}\mbox{He} n \pi{}^{0}$ and $dd\rightarrow$ $^{3}\hspace{-0.03cm}\mbox{He} p \pi{}^{-}$ near the $\eta$ production threshold. This report includes the description of the experimental method and the status of the data analysis.

%~The beam momentum was varying continuously from 2.127 GeV/c to 2.422 GeV/c corresponding to the excess energy range \mbox{Q$\in$(-70,30)~MeV}. 
}

\maketitle

\vspace{-0.3cm}

\section{Introduction}
\label{intro}

The existence of $\eta$-mesic nuclei in which the $\eta$ meson is bound within a nucleus via the strong
interaction was postulated over 20 years ago by Haider and Liu \cite{HaiderLiu1}. Since then$\eta$- and $\eta'$-mesic bound states heave been searched for in many laborathories as e.g.: COSY~\cite{MoskalSmyrski, MSkurzok, Adlarson_2013, Budzanowski, Mersmann, Smyrski1, Krzemien1}, ELSA~\cite{Nanova1,Nanova2}, GSI~\cite{Tanaka}, JINR~\cite{Afanasiev}, JPARC~\cite{Fujioka}, LPI~\cite{Baskov}, and MAMI~\cite{Krusche_2013, Pheron}. Recent theoretical investigations e.g.~\cite{BassTom1, WycechKrzemien, Hirenzaki, Hirenzaki1, Friedman_2013, Wilkin2, Nagahiro_2013, Nagahiro_2012, Kelkar, BassTom} support the search for $\eta$ and $\eta'$-mesic bound states, however, till now none of experiments confirmed its existence. 
The discovery of this new kind of an exotic nuclear matter would be very important for better understanding of the $\eta$ and $\eta'$ meson properties~\cite{InoueOset} and their interaction with nucleons inside nuclear matter. Furthermore it would provide information about the $N^{*}$(1535) resonance~\cite{Jido}, as well as about the flavour singlet component of the quark-gluon wave function of the $\eta$ and $\eta'$ mesons~\cite{BassTom}.\\

%The binding energies of η and η ′ in medium are sensitive to the non-perturbative glue associated with the axial U(1) dynamics [73; 59], and as stated in [119] ”due to the UA(1) anomaly effect, a relatively large mass reduction of η ′ meson is expected at nuclear saturation density, which may indicate the existence of the η ′ -mesic nucleus”. However, so far most of the experimental studies have been concentrated on the search for the η-mesic nuclei because the η-nucleon interaction seems to be much stronger than the η ′ -nucleon or π-nucleon [12]. Yet, recently there are vigourous theoretical [59; 116; 120; 121; 122] and experimental [123; 124; 125] investigations of feasibility of the observation of the η ′ -mesic nuclei started by the predictions published in [126].
\vspace{-0.3cm}

\section{Experimental status}
\label{sec-1}

The search for the $^{4}\hspace{-0.03cm}\mbox{He}$-$\eta$ bound state was carried out with unique accuracy with the WASA facility, installed at the COSY synchrotron in Forschungszentrum J\"ulich. The detection system allows the exclusive
measurement of all ejectiles while continuously changing the beam momentum around the $\eta$ production threshold.  
The signature of the \mbox{$\eta$-mesic} nuclei is searched for via studying the excitation function for the chosen decay channels of the \mbox{$^{4}\hspace{-0.03cm}\mbox{He}$-$\eta$} system, formed in deuteron-deuteron collision~\cite{Moskal1,Krzemien,Moskal_FewBody}.~Till now two experiments were performed focusing on the bound state decay into $^{3}\hspace{-0.03cm}\mbox{He}$ and a nucleon-pion pair~\cite{MSkurzok,WKrzemien_2014}.

The first one, was carried out in June 2008, by measuring the excitation function of the $dd\rightarrow$ $^{3}\hspace{-0.03cm}\mbox{He} p \pi{}^{-}$ reaction near the $\eta$ production threshold covering the excess energy range from -51.4 MeV up to 22 MeV. In the excitation function there is no structure which could be interpreted as a resonance originating from decay of the $\eta$-mesic $^{4}\hspace{-0.03cm}\mbox{He}$. The upper limit for the cross-section for the bound state formation and decay in the \mbox{$dd \rightarrow$ ($^{4}\hspace{-0.03cm}\mbox{He}$-$\eta)_{bound} \rightarrow$ $^{3}\hspace{-0.03cm}\mbox{He} p \pi{}^{-}$} reaction varies from 20 nb to 27 nb at 90\%CL~\cite{Adlarson_2013}. 

\begin{figure}[h]
\centering
\includegraphics[width=5.0cm,height=3.8cm]{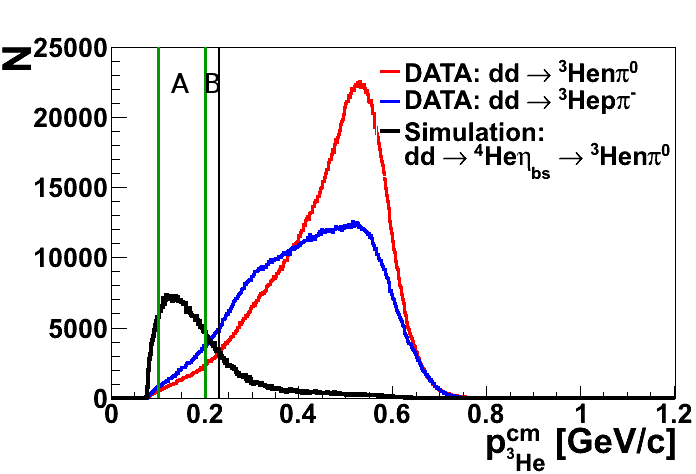} \includegraphics[width=5.0cm,height=3.8cm]{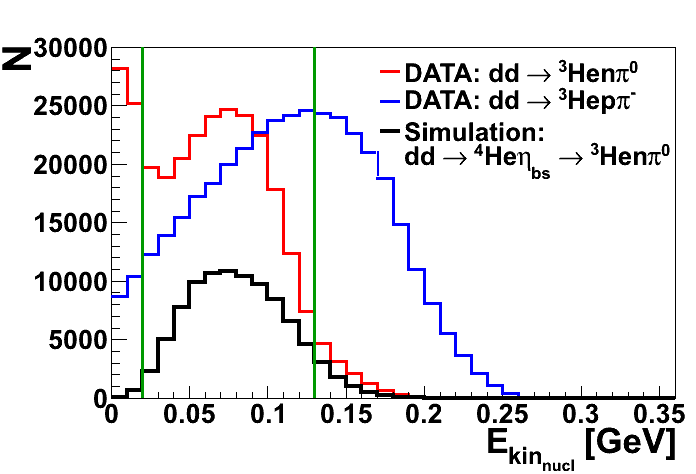}\\
%\vspace{-0.1cm}
\includegraphics[width=5.0cm,height=3.8cm]{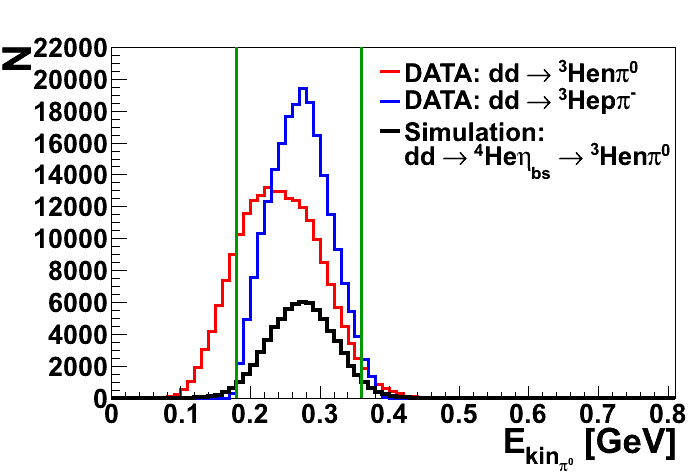} \includegraphics[width=5.0cm,height=3.8cm]{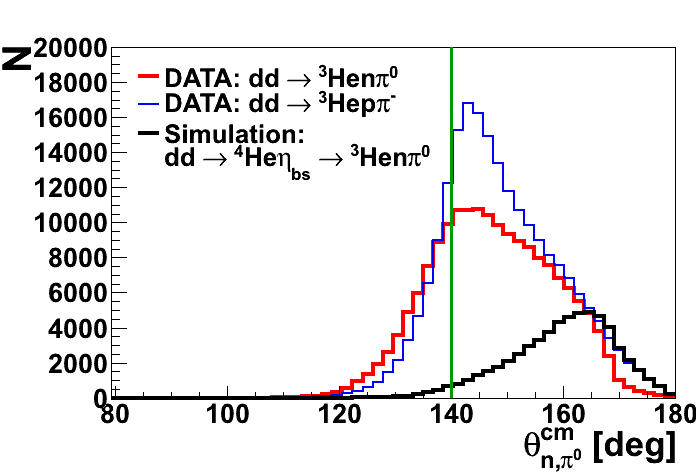}
\vspace{-0.2cm}
\caption{Spectrum of $p^{cm}_{^{3}\hspace{-0.05cm}He}$ (left upper panel), $E^{cm}_{kin_{nucl}}$ (right upper panel), $E^{cm}_{kin_{\pi}}$ (left lower panel) and $\theta^{cm}_{nucl,\pi}$ (right lower panel). Data are shown in red and blue for $dd\rightarrow$ $^{3}\hspace{-0.03cm}\mbox{He} n \pi{}^{0}$ and $dd\rightarrow$ $^{3}\hspace{-0.03cm}\mbox{He} p \pi{}^{-}$ reaction, respectively. Monte Carlo simulations of the signal reaction are shown in black, while the applied cuts are marked bthe green lines.~\label{fig2}}
\end{figure}

In the second experiment, in November 2010, we collected about 20 times higher statistics with respect to the previous measurement~\cite{Adlarson_2013}.
The search for a $^{4}\hspace{-0.03cm}\mbox{He}$-$\eta$ bound state was performed by measuring the excitation function of the $dd\rightarrow$ $^{3}\hspace{-0.03cm}\mbox{He} n \pi{}^{0}\rightarrow$ $^{3}\hspace{-0.03cm}\mbox{He} n \gamma \gamma$ and $dd\rightarrow$ $^{3}\hspace{-0.03cm}\mbox{He} p \pi{}^{-}$ reactions in the vicinity of the $\eta$ production threshold. The deuteron beam momentum was varying continuously within each acceleration cycle from 2.127 GeV/c to 2.422 GeV/c which corresponds to a range of excess energies \mbox{Q$\in$(-70,30)~MeV}. 
Independent analyses for the $dd\rightarrow$ $^{3}\hspace{-0.03cm}\mbox{He} n \pi{}^{0}\rightarrow$ $^{3}\hspace{-0.03cm}\mbox{He} n \gamma \gamma$ and $dd\rightarrow$ $^{3}\hspace{-0.03cm}\mbox{He} p \pi{}^{-}$ reactions were carried out. The $^{3}\hspace{-0.03cm}\mbox{He}$ for both cases was identified in the Forward Detector based on the \mbox{$\Delta$E-E method}. The neutral pion $\pi^{0}$ was reconstructed in the Central Detector from the invariant mass of two gamma quanta originating from its decay while the $\pi^{-}$ identification in the Central Detector was based on the measurement of the energy loss in the Plastic Scintillator combined with the energy deposited in the Electromagnetic Calorimeter. Neutron and proton were identified via the missing mass technique. 
%The appropriate spectra with applied cuts are presented in Fig.~\ref{fig1}.
Events corresponding to the bound states production were selected via applying appropriate cuts in the $^{3}\hspace{-0.03cm}\mbox{He}$ center of mass (CM) momentum, nucleon CM kinetic energy, pion CM kinetic energy and the opening angle between nucleon-pion pair in the CM based on Monte Carlo simulations. These cuts are presented in Fig.~\ref{fig2}. 

\vspace{-0.2cm}

The excitation functions for both reactions were determined for a "signal rich" region corresponding to momenta of the $^{3}\hspace{-0.03cm}\mbox{He}$ in the CM system with \mbox{$p^{cm}_{^{3}\hspace{-0.05cm}He}\in(0.1,0.2)$GeV/c} and for a "signal poor" region with \mbox{$p^{cm}_{^{3}\hspace{-0.05cm}He}\in(0.2,0.23)$GeV/c} which are marked with (A) and (B) in the left upper panel in Fig.~\ref{fig2}, respectively.

\begin{figure}[h]
\centering
\includegraphics[width=6.0cm,height=5.0cm]{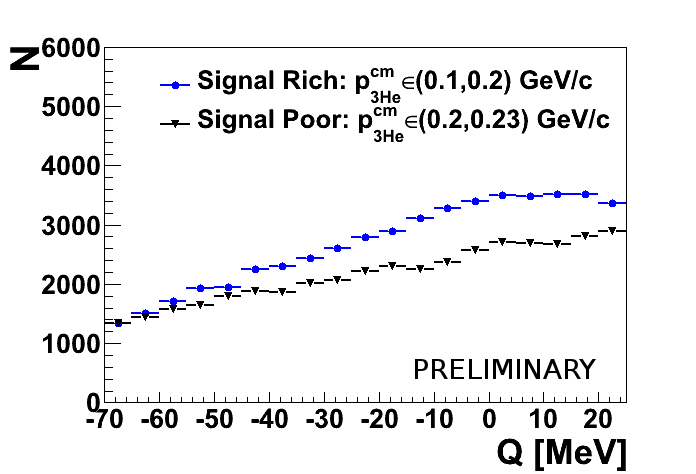}
\includegraphics[width=6.0cm,height=5.0cm]{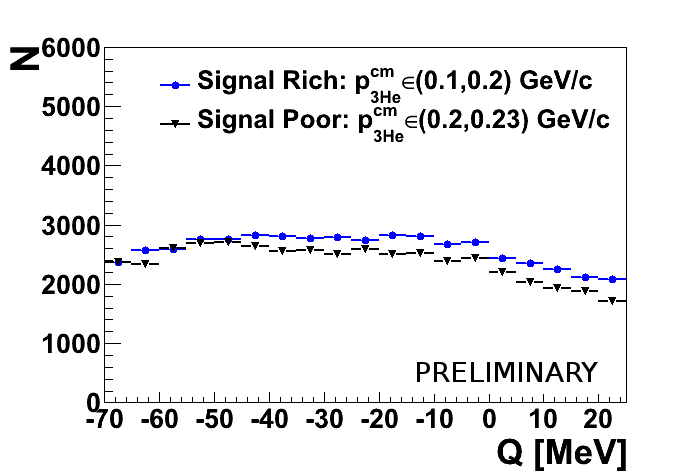}
%\vspace{-0.3cm} 
\caption{Excitation function for the $dd\rightarrow$ $^{3}\hspace{-0.03cm}\mbox{He} n \pi{}^{0}$ reaction (left panel) and  the \mbox{$dd\rightarrow$ $^{3}\hspace{-0.03cm}\mbox{He} p \pi{}^{-}$} reaction (right panel). The "signal rich" region is shown in blue while the "signal poor" region normalized to the first bin of the "signal rich" region is shown in black. The analysis is based on the whole data sample.
\label{fig3}}  
\end{figure}

The excitation functions for the \mbox{$dd\rightarrow$ $^{3}\hspace{-0.03cm}\mbox{He} n \pi{}^{0}$} and \mbox{$dd\rightarrow$ $^{3}\hspace{-0.03cm}\mbox{He} p \pi{}^{-}$} reactions are presented in Fig.~\ref{fig3}. 
The preliminary results presented in Fig.~\ref{fig3} reveal no structure which could be interpreted as a signature of a bound state. The shapes of the spectra in the "signal rich" and "signal poor" regions differ from each other. This is possibly a promissing result, but its final interpretation requires detailed simulations in order to understand the background contributions to the observed excitation functions. The analysis is still in progress. However, already now we can conclude that a collected data are of a very good quality and that we have achieved a sensitivity of the cross section of the order of few nb for the bound state production in $dd \rightarrow$ ($^{4}\hspace{-0.03cm}\mbox{He}$-$\eta)_{bound} \rightarrow$ $^{3}\hspace{-0.03cm}\mbox{He} n \pi{}^{0}$ and $dd \rightarrow$ ($^{4}\hspace{-0.03cm}\mbox{He}$-$\eta)_{bound} \rightarrow$ $^{3}\hspace{-0.03cm}\mbox{He} p \pi{}^{-}$ reactions.

%\vspace{-0.3cm}

%\begin{figure}[h]
%\centering
%\includegraphics[width=5.0cm,height=3.8cm]{IM_pi0_lev2_cut1_DATA_Sig_publ.png} 
%\includegraphics[width=5.0cm,height=3.8cm]{Edep_CalvsPSB_lev2_cut1_publ.png}\\
%\includegraphics[width=5.0cm,height=3.8cm]{MM_neutron_lev2_cut2_DATA_Sig_mx_Ex_publ.png} 
%\includegraphics[width=5.0cm,height=3.8cm]{MM_proton_lev2_cut2_DATA_Sig_Woj_publ.png}
%\vspace{-0.3cm}
%\caption{$\pi^{0}$ and $\pi^{-}$ identification (upper panel). Proton and neutron identification (lower panel). The data were marked with red and blue lines, the Monte Carlo simulations of the signal are marked with the black line, while the applied cuts are marked in green.~\label{fig1}}
%\end{figure}

\vspace{-0.3cm}

%\newpage
\section{Perspectives}

The recent theoretical and experimental results~\cite{Friedman_2013, Wilkin_Acta2014, Gal_2014, Kelkar, Nagahiro_2013, Nagahiro_2012, Krusche_2013, Krusche_2014} bring new and stronger arguments in favour of carrying out the search for the $^{3}\hspace{-0.03cm}\mbox{He}$-$\eta$ bound state. Therefore, in May 2014, we extended the measurement to search for the $\eta$-mesic helium for processes corresponding to two anticipated mechanisms: (i) absorption of the $\eta$ meson by one of the nucleons, which subsequently decays into $N^{*}$ - $\pi$ pair e.g.: $pd \rightarrow$ ($^{3}\hspace{-0.03cm}\mbox{He}$-$\eta)_{bound} \rightarrow$ $p p p \pi{}^{-}$ , and (ii) decay of the $\eta$ -meson while it is still "orbiting" around a nucleus e.g.: $pd \rightarrow$ ($^{3}\hspace{-0.03cm}\mbox{He}$-$\eta)_{bound} \rightarrow$ $^{3}\hspace{-0.03cm}\mbox{He} 6\gamma$ or $pd \rightarrow$ ($^{3}\hspace{-0.03cm}\mbox{He}$-$\eta)_{bound} \rightarrow$ $^{3}\hspace{-0.03cm}\mbox{He} 2\gamma$ reactions. Almost two weeks of measurement with an average luminosity of about 6$\cdot10^{30} cm^{-2} s^{-1}$ allowed to collect high statistics of data. The analysis is in progress.
In the first case we expect a cross section for the bound state formation in the order of 80 nb. In the case of the second considered mechanism, we may roughly estimate the cross section to be about 0.4 nb taking into account that the total width of the $\eta$ meson is about 1.3 keV, the width of the $^{3}\hspace{-0.03cm}\mbox{He}$-$\eta$ is less than about 500 keV, and the 2$\gamma$ and 6$\gamma$ branching ratios amounts to about 39\% and 33\%, respectively~\cite{Wilkin_Acta2014}.

\begin{acknowledgement}

\noindent We acknowledge support by the Foundation for Polish Science - MPD program, co-financed by the European
Union within the European Regional Development Fund, by the Polish National Science Center through grant No. 2011/01/B/ST2/00431 and by the FFE grants of the Research Center J\"ulich.

\end{acknowledgement}
%
% BibTeX or Biber users please use (the style is already called in the class, ensure that the "woc.bst" style is in your local directory)
% \bibliography{name or your bibliography database}
%
% Non-BibTeX users please use

\end{document}